\documentclass[aps,prb,twocolumn,showpacs,showkeys]{revtex4-1}

\usepackage{epsfig}
\usepackage{amssymb}

\usepackage{graphicx}
\usepackage{dcolumn}
\usepackage{bm}
\usepackage{color}
\usepackage{hyperref}
\begin{document}

\title{Spin Torque in Anisotropic Tunneling Junctions}
\date{\today}
\author{A. Manchon$^{1}$}
\affiliation{$^{1}$Division of Physical Science and Engineering, Materials Science and Eng., KAUST, Thuwal 23955-6900, Saudi Arabia.}
\begin{abstract}
Spin transport in magnetic tunnel junctions comprising a single magnetic layer in the presence of interfacial spin-orbit interaction (SOI) is investigated theoretically. Due to the presence of interfacial SOI, a current-driven spin torque can be generated at the second order in SOI, even in the absence of an external spin polarizer. This torque possesses two components, in-plane and perpendicular to the plane of rotation, that can induce either current-driven {\em magnetization switching} from in-plane to out-of-plane configuration or {\em magnetization precessions}, similarly to Spin Transfer Torque in spin-valves. Consequently, it appears that it is possible to control the magnetization steady state and dynamics by either varying the bias voltage or electrically modifying the SOI at the interface.
\end{abstract}
\maketitle

Since the theoretical prediction and experimental observation of current-driven magnetization control \cite{slonc,tsoi}, spintronics is considered as a promising candidate for low energy consumption devices \cite{katine}. Up until now, Spin Transfer Torque (STT) has essentially been observed in inhomogeneous magnetic structures such as Spin-Valves, Magnetic Tunnel Junctions and Magnetic Domain Walls \cite{review}. Moreover, recent experimental observations on single ferromagnet-based structures have opened promising opportunities for new device concepts. For example, Tunneling Anisotropic Magnetoresistance (TAMR) has been observed in Magnetic Tunnel Junctions (MTJs) comprising a single ferromagnet \cite{gould,park,uemura,moser} (referred to as Semi-MTJs - SMTJs). Ref. \onlinecite{manchon} proposed to exploit the spin-orbit interaction (SOI) present in a single ferromagnetic layer to electrically control the magnetization direction \cite{miron,leonid}. Alternatively, the voltage-controlled manipulation of magnetic anisotropy of thin magnetic layers through thick insulators has been achieved \cite{nano}. Efficient electrical control of the magnetization direction of a single ferromagnetic layer combined with sizable TAMR effect would offer powerful perspectives for spin-based memory devices.\par

The key ingredient of TAMR \cite{gould,park,uemura,moser}, SOI-induced torque (SOI-ST) \cite{manchon} and magnetic anisotropy \cite{nano} is the SOI arising either from Bulk Inversion Asymmetry \cite{leonid} (BIA) or Structure Inversion Asymmetry \cite{miron,nano} (SIA). In the latter case, the presence of a large potential gradient ${\bm \nabla} V$ at the interface between, say, an insulator and a metal, generates a local electric field perpendicular to the interface ${\bm \nabla} V=-E{\bf z}$, inducing a SOI of the form \cite{rashba}
\begin{equation}\label{eq:rashba}
{\hat H}_R({\bf k})={\bf B}_{R}({\bf k})\cdot{\bm{\hat \sigma}}=\alpha_R({\bf k}\times{\bf z})\cdot{\bm{\hat \sigma}}.
\end{equation}
The precise form and magnitude of $\alpha_R$ terms has been widely studied in semiconductor 2DEG \cite{rashba,nitta}, and only recently at metallic interfaces and can be as large as 1-5eV.\AA$^2$ \cite{metals,moser}.

In this Letter, we suggest that the presence of such an interfacial SOI at the interface between a ferromagnet and an insulator in a SMTJ is responsible for a non-equilibrium spin torque. We demonstrate that the magnetization of the ferromagnetic layer of such a junction can be controlled or excited by an external bias voltage applied across the junction. In the non-equilibrium regime, the interfacial SOI generates a bias voltage-driven spin torque on the ferromagnetic layer inducing either {\em magnetization switching} or {\em self-sustained magnetic precessions}. This effect belongs to the family of SOI-induced spin torques \cite{manchon,soitorque} and the correspondences with the conventional STT observed in spin-valves \cite{slonc,tsoi} will be discussed at the end of this paper.\par
\begin{figure}[!h]
	\centering
	\begin{tabular}{cc}
	\begin{tabular}{c}
		\includegraphics[width=4cm]{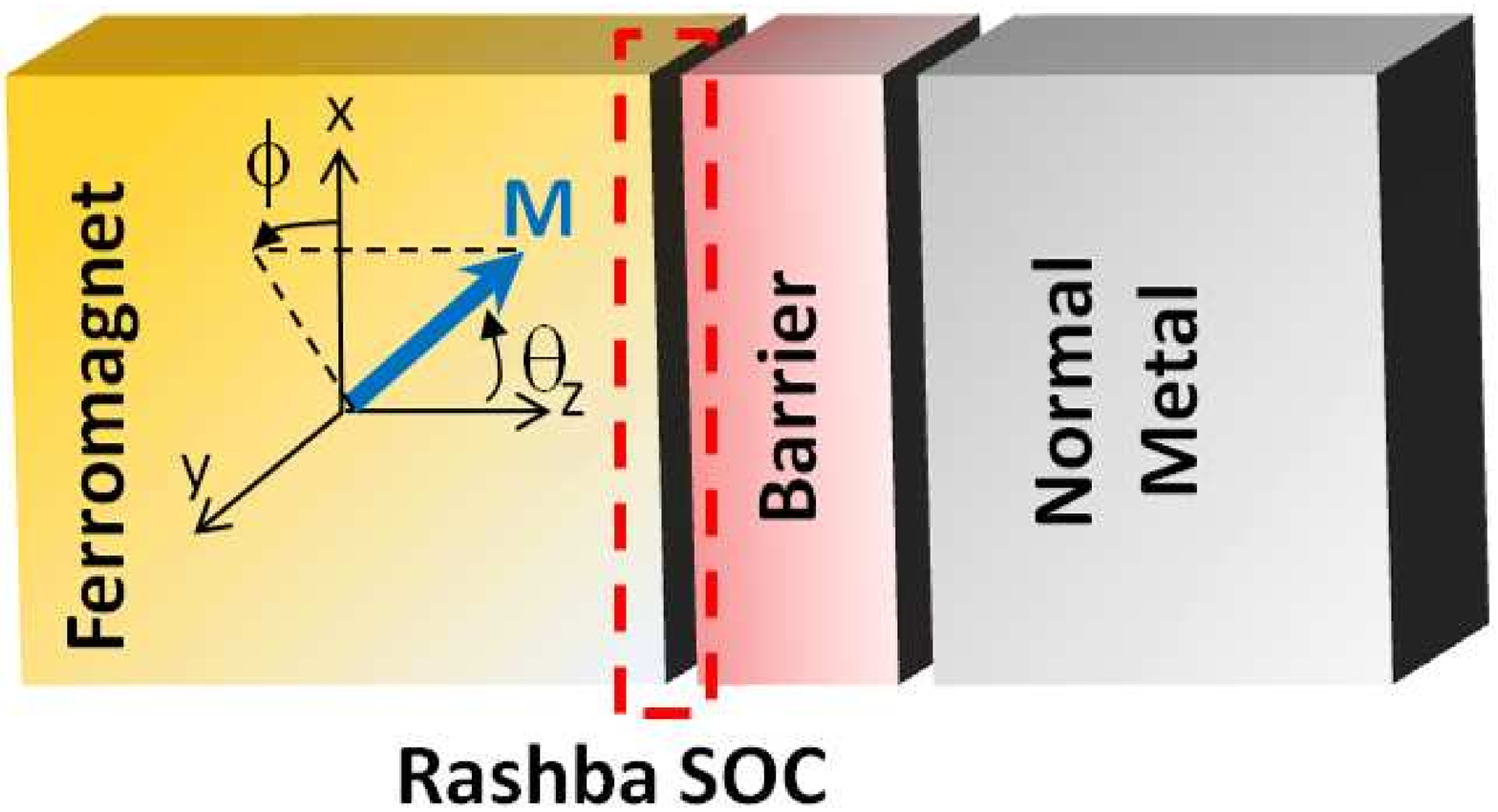}\\
		\includegraphics[width=3cm]{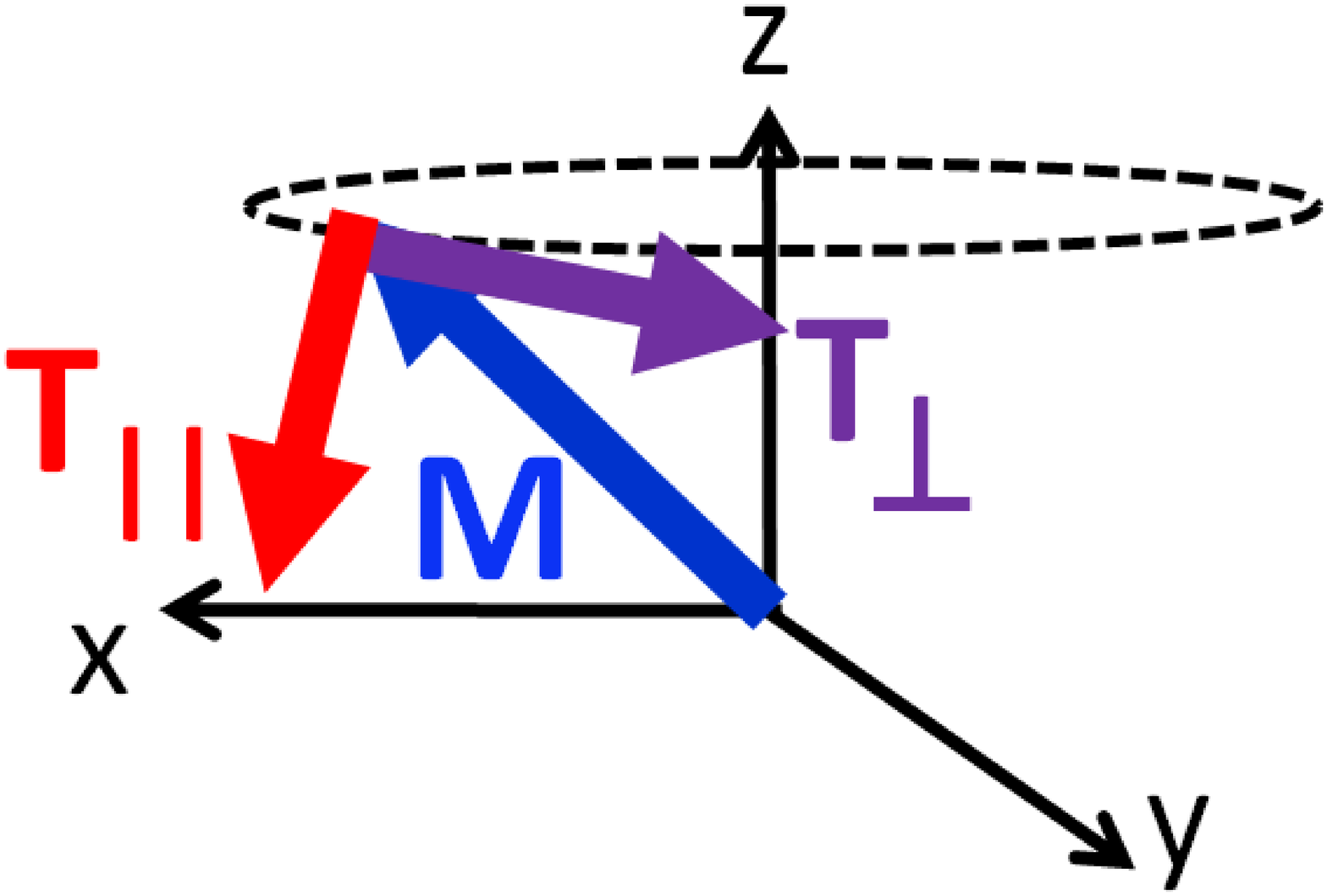}
		\end{tabular}&
		\begin{tabular}{cc}
	\includegraphics[width=4cm]{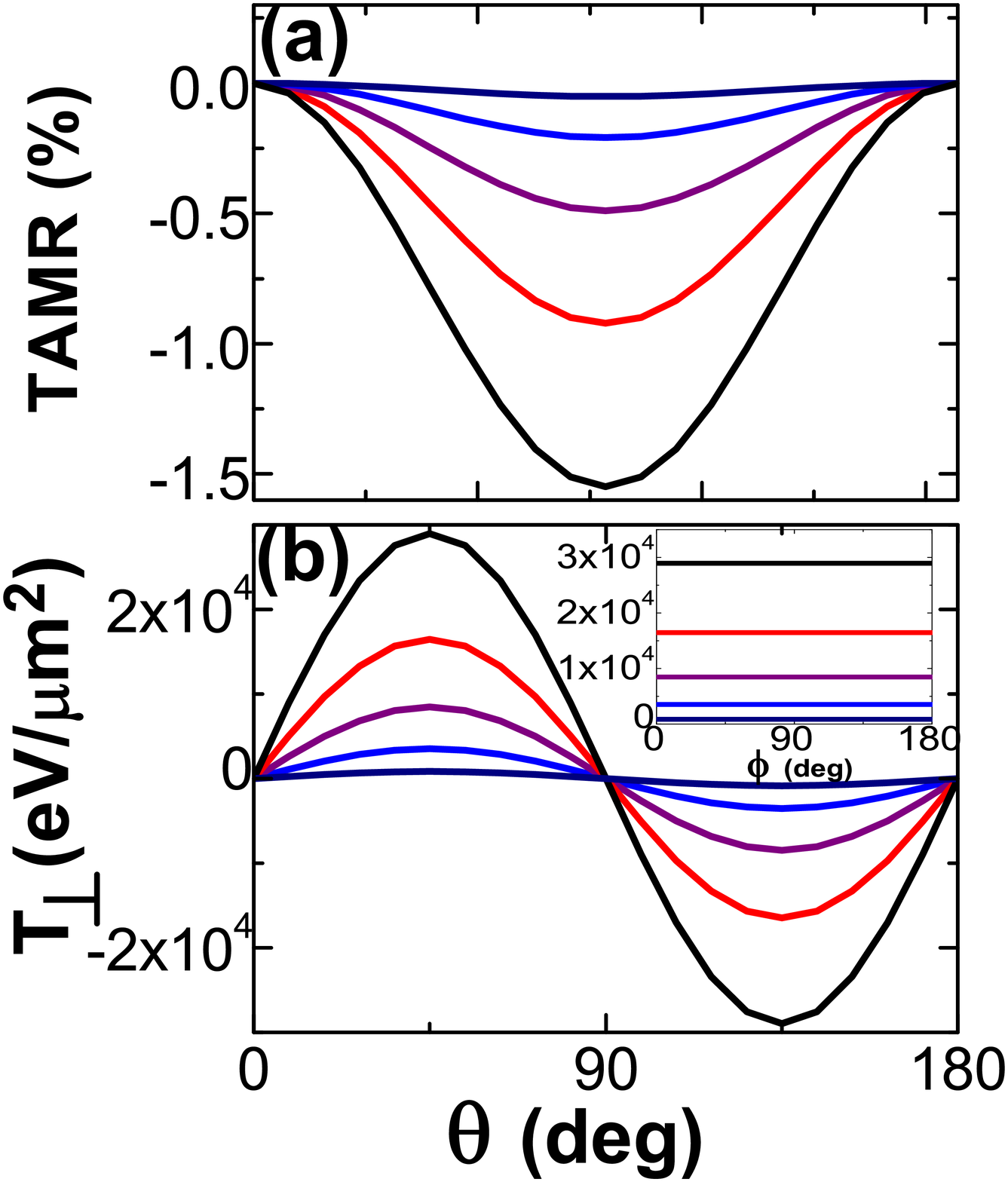}
		\end{tabular}
		\end{tabular}
	\caption{\label{fig:Fig1}(Color online) Left panel: [top] Schematics of a Semi-Magnetic Tunnel Junction. [bottom] Influence of the spin torque on the magnetization; Right panel: Angular dependence of (a) TAMR and (b) perpendicular torque $T_{\bot}$ at zero bias, for a rotation in the (010) planes and $\alpha_R\in$[1,5]eV.\AA$^2$. The parameters are adapted for Fe/MgO interface: $U_0$=1eV, d=0.6nm, $k_F^\uparrow=1.09$nm$^{-1}$, $k_F^\downarrow=0.4$nm$^{-1}$. Inset: $\phi$-dependence of the perpendicular torque.}
\end{figure}

In regular MTJs, the transport is essentially controlled by the interfacial densities of states \cite{mtj}. Therefore, the tunneling magnetoresistance (TMR) and STT are proportional to the interfacial polarization of the electrodes, $P_i$: $TMR\propto P_LP_R$, $\tau_{||}\propto P_L\sin\theta$, and $\tau_{\bot}\propto P_LP_R\sin\theta$ \cite{mtj}. In contrast, in the case of SMTJs, $P_L=0$ and both STT and TMR should vanish. However, in the presence of interfacial SOI it has been found that the resistance of the junction depends on the orientation of the magnetization against the interface \cite{gould}.\par

This TAMR effect has been observed using semiconductors \cite{gould}, metals \cite{park} and hybrid structures \cite{uemura,moser}. Theoretical investigations \cite{moser,matos} have shown that TAMR is well described with Rashba and Dresselhaus effective SOI \cite{rashba,tsymbal2} (the latter being restricted to non-centrosymmetric crystals). Following Eq. (\ref{eq:rashba}), Rashba SOI induces an angular momentum transfer from the spin momentum to the orbital degree of freedom. Since the interfacial SOI is linear in ${\bf k}$, this effect vanishes at the first order in SOI after averaging over the Fermi sphere [$\langle{\hat H}_{R}({\bf k})\rangle=0$]. But in a ferromagnet, due to the presence of the local magnetic exchange, a net transfer of angular momentum occurs at the second order [$\langle{\hat H}_R({\bf k})^2\rangle\neq0$]. As a consequence, magnetic anisotropy as well as TAMR appear, proportional to the second order in Rashba SOI, $\propto\alpha_R^2$ \cite{matos,tsymbal2}. Interestingly, since the angular momentum transfer involves only {\em in-plane components} of the spin and orbital momentum (i.e. $\sigma_{x,y}$ and $k_{x,y}$), the itinerant spin density produced by this transfer lies in the (x,y) plane only. 

In the present work, we consider the F/I/N trilayer depicted in Fig. \ref{fig:Fig1}, where F is a ferromagnetic layer (Co, Fe, Ni and compounds), I is an insulator (MgO, AlOx, GaAs) and N is a normal metal (Cu, Ag, Au, Pt etc.). In order to capture the most relevant features of the mechanism described here, we choose a minimal model only considering the most pertinent material parameters. Matos-Abiague et al. \cite{matos} showed that in the case of centrosymmetric barriers (such as AlOx, MgO), the TAMR is mostly due to the Rashba SOI at the interface between the ferromagnet and the tunnel barrier. The free electron Hamiltonian of the junction is then
\begin{eqnarray}{\hat H}&=&-\frac{\hbar^2}{2}\nabla\frac{1}{m(z)}\nabla+U(z)+{\hat H}_{R}\delta(z-z_L).
\end{eqnarray}
${\hat H}_{R}$ is given in Eq. (\ref{eq:rashba}), $m(z)$ is the effective mass of the electron, equal to $m_0$ in the electrodes and $m_{eff}m_0$ in the barrier. $U(z)$ is the potential of the junction, given by
\begin{eqnarray}
&&U_{z<0}=J{\bm{\hat \sigma}}\cdot{\bf M}+\frac{eV_b}{2},\;U_{z>d}=-\frac{eV_b}{2}\\
&&U_{0<z<d}=U_0+(\frac{1}{2}-\frac{z}{d})eV_b,
\end{eqnarray}
where $U_0$ and $d$ are the barrier height and thickness, $V_b$ is the bias voltage, $J$ is the $s-d$ exchange coupling, ${\bm{\hat \sigma}}$ is the vector of Pauli spin matrices and ${\bf M}=(\sin\theta\cos\phi,\sin\theta\sin\phi,\cos\theta)$ is the magnetization direction of the ferromagnetic electrode [see Fig. \ref{fig:Fig1}].\par

The mechanism giving rise to itinerant spin density can be understood by looking at the spin density continuity equation $\frac{d{\bf m}}{dt}=\frac{1}{i\hbar}\langle[{\bm{\hat \sigma}},{\hat H}]\rangle$, which reads
\begin{eqnarray}
\frac{d{\bf m}}{dt}=\frac{i\hbar}{m}{\bm \nabla}\cdot\langle{\bm{\hat \sigma}}\otimes{\bm \nabla}\rangle-\frac{2J}{\hbar}{\bf m}\times{\bf M}+\frac{1}{i\hbar}\langle[{\bm{\hat \sigma}},{\hat H}_R]\rangle.
\end{eqnarray}
The first term is the regular spin current divergence in the absence of SOI, the second term is the torque exerted by the itinerant spin density ${\bf m}$ on the local magnetization ${\bf M}$ and the last term is the torque between the itinerant spin and orbital angular momentum. For instance, in MTJs the last term is generally zero and the spin torque is directly associated with the spatial variation of the spin current $\propto\langle{\bm{\hat \sigma}}\otimes{\bf \nabla}\rangle$ \cite{mtj}. In the present case, Rashba SOI acts like a source for spin density and therefore the spin torque is no more simply related to the spin current. Consequently, the proper way to evaluate the spin torque is to calculate directly the local itinerant spin density ${\bf m}(z)$. We assume a semi-infinite magnetic layer, as is usually done in magnetic tunnel junctions \cite{mtj}: The total spin torque exerted by the transverse spin density ${\bf m}$ on the magnetization ${\bf M}$ is defined as ${\bf T}=J\int_V {\bf m}\times{\bf M}dV$, where V is the volume of the magnetic layer.\par

The charge and spin currents are then evaluated using the conventional definitions
\begin{eqnarray}
J_e&=&\frac{e}{\hbar}\Im[\sum_{s,i}\int dEd^2{\bf k}_{||}\Psi^{s*}_i(E,{\bf k}_{||})\partial_z\Psi^s_i(E,{\bf k}_{||})f_i],\\
{\bf m}&=&\sum_{s,i}\int dEd^2{\bf k}_{||}\Psi^{s*}_i(E,{\bf k}_{||}){\bm{\hat \sigma}}\Psi^s_i(E,{\bf k}_{||})f_i,\label{eq:0}
\end{eqnarray}
$\Psi^{s}_i(E,{\bf k}_{||})$ being the Hartree-Fock two-component wave function for an electron of energy $E$ and in-plane wave vector ${\bf k}_{||}$, issued from the $i$-th reservoir with a Fermi-Dirac distribution $f_i$. Assuming convenient material parameters (barrier characteristics and effective mass), the present minimal model satisfyingly reproduces the TAMR obtained by Matos-Abiague et al. \cite{moser,matos} in the case of Fe/GaAs/Au and the one obtained by Park et al. \cite{park} in (Pt/Co)$_n$/AlOx/Pt, with a reasonable degree of accuracy. The angular dependence of the zero bias conductance for rotation of the magnetization in the (010) plane for $\alpha_R=1-5$eV.\AA$^2$ is displayed in Fig. \ref{fig:Fig1}(a) and shows the expected $(\cos2\theta-1)$-dependence \cite{moser,matos}. This angular dependence can be understood by noticing that the junction is physically equivalent upon the transformation $\theta\rightarrow-\theta\rightarrow\theta+\pi$ \cite{matos}.

Since the torque is by definition perpendicular to the magnetization, it reads
\begin{eqnarray}\label{eq:b}
&&{\bf T}=T_{||}{\bf M}\times({\bf z}\times{\bf M})+T_{\bot}{\bf z}\times{\bf M}.
\end{eqnarray}
The spin torque possesses two components that can be referred to as in-plane $T_{||}$, and perpendicular torques, $T_\bot$. At zero bias, only the perpendicular torque $T_{\bot}$ is non-zero and displays an angular dependence on the form $\sin2\theta$ [Fig. \ref{fig:Fig1}(b)] which favors the {\em perpendicular} configuration. As mentioned above, the itinerant spin density due to Rashba SOI lies in the (x,y) plane of the junction [Eq. (\ref{eq:rashba})]. Consequently (i) when the magnetization is oriented along {\bf z} ($\theta=0$), no transfer occur ($\langle\sigma_{x,y}\rangle=0$) and the spin torque vanishes; (ii) when the magnetization is oriented in the (x,y) plane ($\theta=\pi/2$), since the spin density lies itself in this plane, the spin torque is also zero. Finally, the junction is invariant under $\phi$-rotation, so the spin torque (and resistance) does not depend on $\phi$ [Fig. \ref{fig:Fig1}(b), inset]. This gives rise to an angular dependence on the form $\sin2\theta$ as well as four associated stable magnetic states: two perpendicular to the plane ($\theta=0,\pi$) and two in the plane ($\theta=\pm\pi/2$).\par
\begin{figure}[!ht]
	\centering
		\includegraphics[width=8cm]{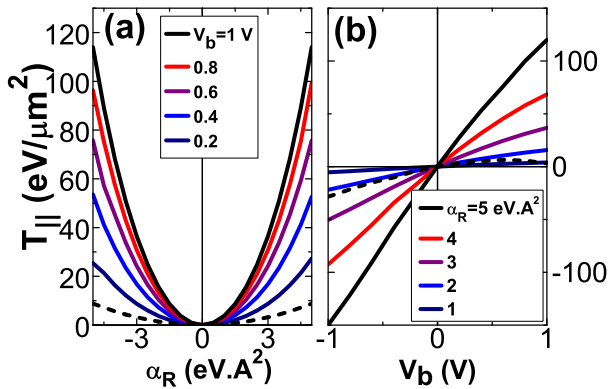}
	\caption{\label{fig:Fig2}(Color online) Non-equilibrium in-plane torque  $T_{||}(V_b)$ as a function of (a) Rashba parameter and (b) bias voltage. The dashed line shows the non-equilibrium perpendicular torque for $V_b$=0.8V (a) and $a_R=4$eV.\AA$^2$ (b). The parameters are the same as in Fig. \ref{fig:Fig1}.}
\end{figure}
When applying a bias voltage across the junction, spin polarized electrons tunnel through the barrier, significantly modifying the spin imbalance in the left electrode. Therefore, both the TAMR and the magnetic anisotropy are strongly affected by the voltage. In Fig. \ref{fig:Fig2} the non-equilibrium in-plane torque, $T_{||}(V_b)$, is represented a function of $\alpha_R$ [Fig. \ref{fig:Fig2}(a)] for different bias voltages, and as a function of the voltage [Fig. \ref{fig:Fig2}(b)] for different $\alpha_R$. The spin torque displays a quadratic dependence on the Rashba parameter as expected from the symmetry of the Rashba Hamiltonian (linear in ${\bf k}$). For larger Rashba parameters, higher orders in $\alpha_R$ appear (not shown). Interestingly, the amplitude of the non-equilibrium perpendicular torque $T_{\bot}(V_b)-T_{\bot}(0)$ is about one order of magnitude smaller than the in-plane torque [dashed line in Fig. \ref{fig:Fig2}(a) and (b)].

The form of the torque displayed in Eq. (\ref{eq:b}) is similar to the usual STT in a MTJ whose polarizer is oriented along ${\bf z}$ \cite{mtj}. Whereas the perpendicular torque $\propto{\bf z}\times {\bf M}$ competes with the demagnetizing field and the perpendicular anisotropy, the in-plane torque $\propto{\bf M}\times({\bf z}\times {\bf M})$ competes with the damping. As a consequence, one expects the current-driven torque to produce current-induced magnetization switching from out-of-plane to in-plane and vice-versa as well as current-driven magnetization precessions.\par
 
\begin{figure}[!ht]
	\centering
		\includegraphics[width=8cm]{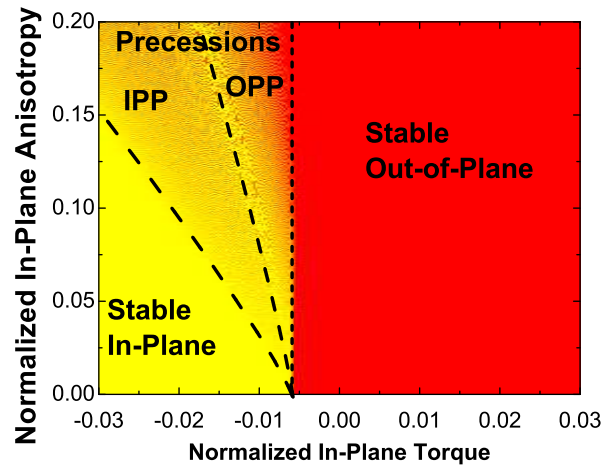}
	\caption{\label{fig:Fig3}(Color online) Stability Diagram of the magnetization in the presence of perpendicular and in-plane anisotropy, as well as in-plane torque. The parameters are magnetic damping $\eta=0.01$, perpendicular anisotropy $Q_{\bot}=0.75$, saturation magnetization $M_s=1.6$T \cite{Ikeda}. The dashed lines are guides for the eye.}
\end{figure}

In order to illustrate the current-driven magnetization dynamics that can be generated in such structures, we numerically solve the macrospin Landau-Lifshitz-Gilbert equation in the presence of spin torque using a fourth-order Runge-Kutta method \cite{review}. A ferromagnetic layer is considered, possessing both in-plane and out-of-plane magnetic anisotropies $Q_{||}=H^K_{||}/M_s$ and $Q_{\bot}=H^K_{\bot}/M_s$, where $H^K_{||}$ ($H^K_{\bot}$) is the in-plane (perpendicular) anisotropy field and $M_s$ is the saturation magnetization. Both anisotropies are needed in order to achieve stable in-plane and out-of-plane magnetization states. Since $T_{\bot}<<T_{||}$, the non-equilibrium perpendicular torque will be disregarded. In addition, {\em no external field} is applied and only the in-plane torque $\tau_{||}=T_{||}/(\mu_0M_s^2d)$ is considered. Finally, for the numerical simulations, we adopt parameters close to the one measured by Ikeda et al. \cite{Ikeda} (see Fig. \ref{fig:Fig3}).\par

\begin{figure}[!ht]
	\centering
	\begin{tabular}{cc}
	\begin{tabular}{c}
		\includegraphics[width=4cm]{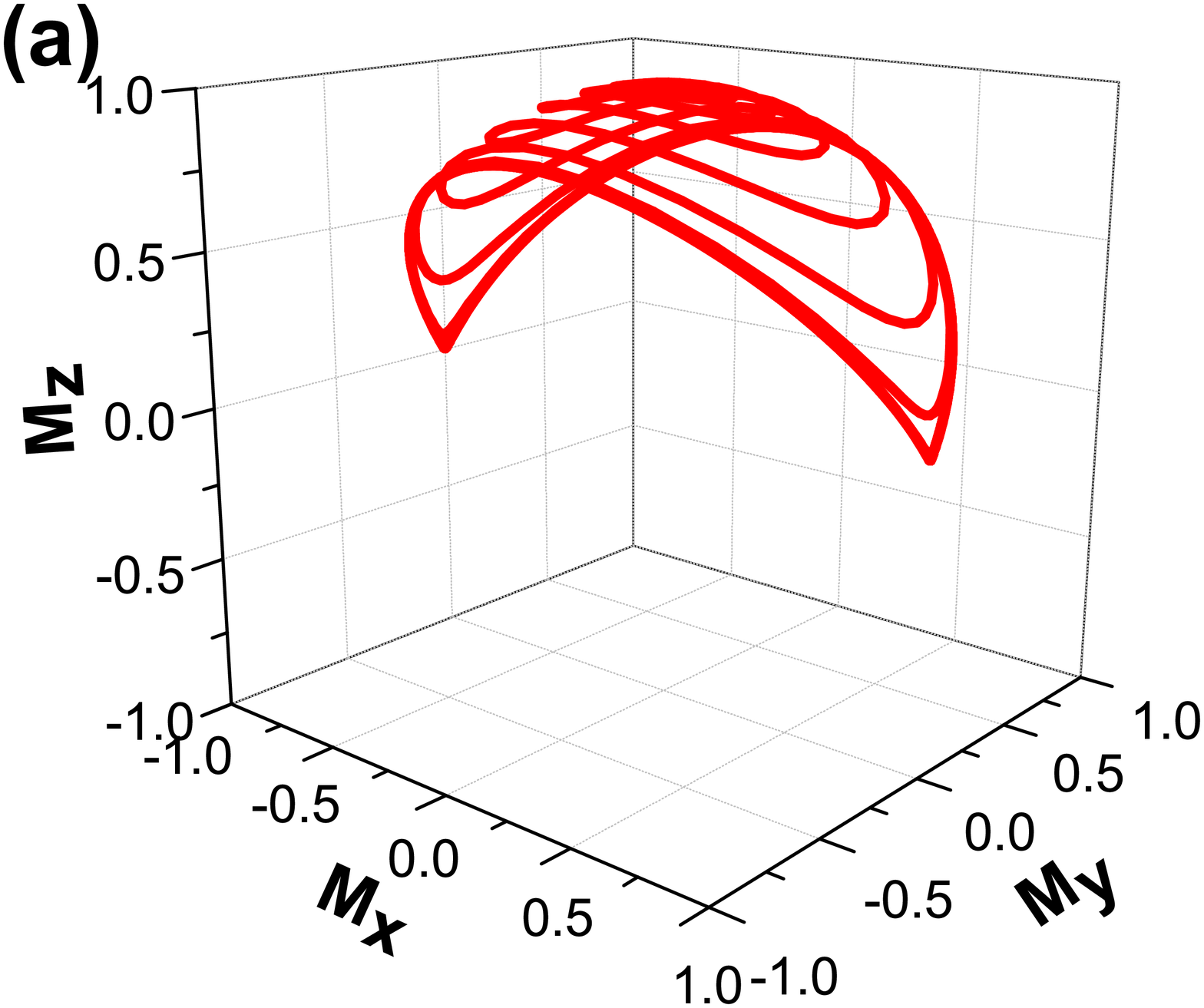}\\
		\includegraphics[width=4cm]{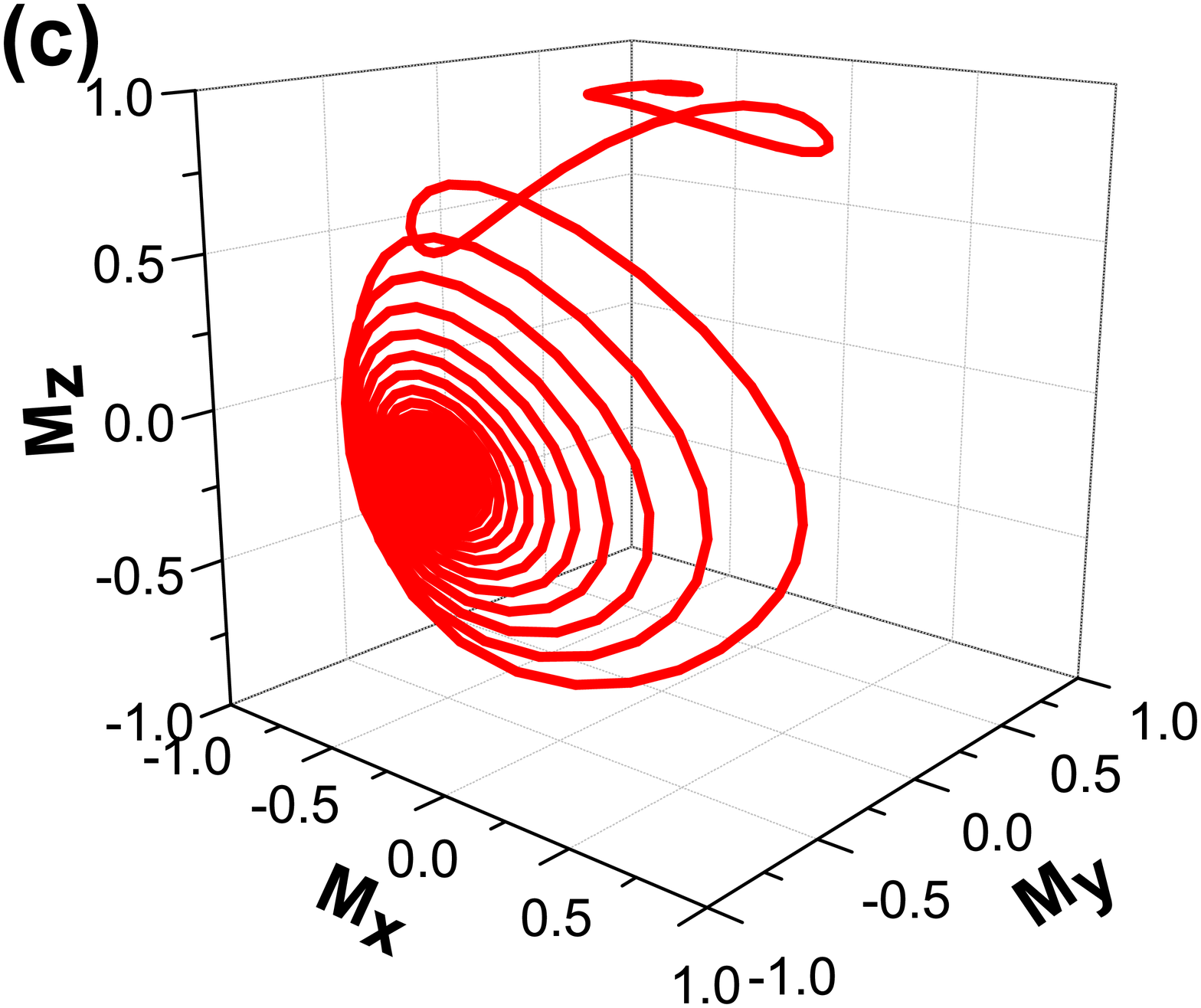}
		\end{tabular}&
		\begin{tabular}{cc}
	\includegraphics[width=4cm]{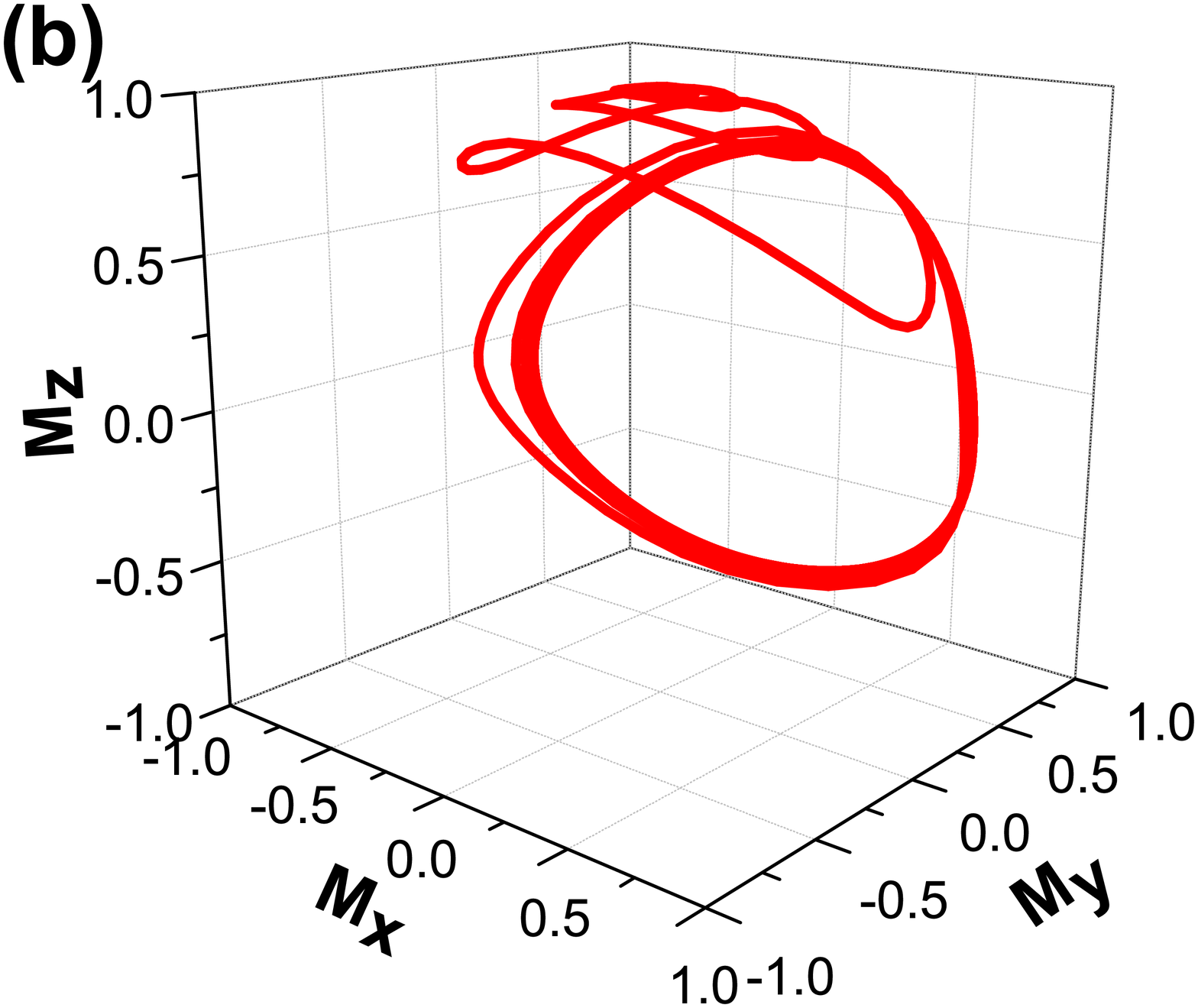}\\
	\includegraphics[width=4cm]{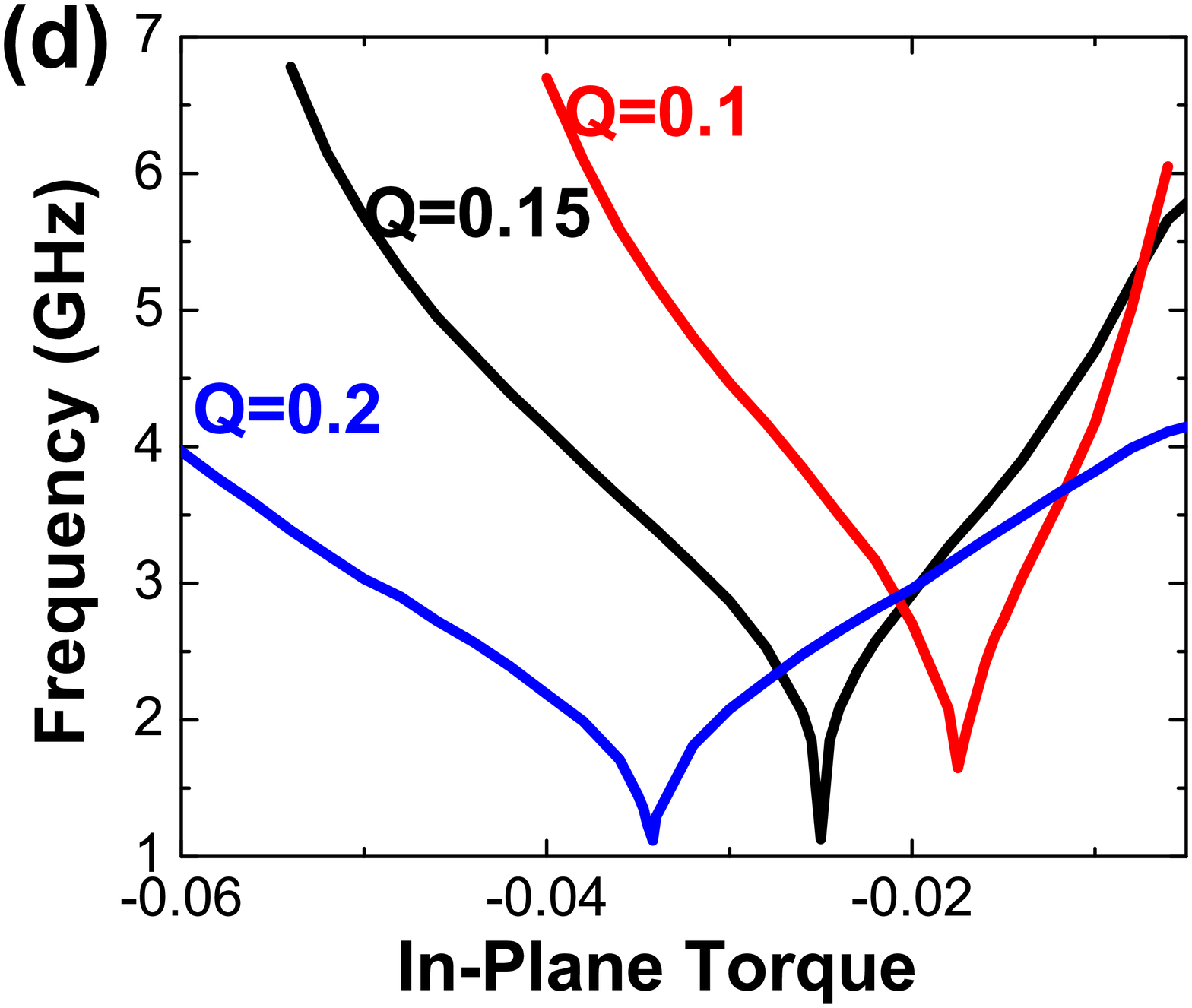}
		\end{tabular}
		\end{tabular}
	\caption{\label{fig:Fig4}(Color online) Magnetization trajectory for a magnetization initially perpendicular to the plane and submitted to an in-plane anisotropy $Q_{||}=0.2$ and (a) $\tau_{||}=-0.0125$, (b) $\tau_{||}=-0.02$, (c) $\tau_{||}=-0.0375$; (d) Dependence of the oscillation frequency as a function of the in-plane torque.}
\end{figure}

Fig. \ref{fig:Fig3} displays the normalized resistance of the junction ($\propto\cos2\theta$) taken at a time t=35ns for a magnetization initially perpendicular to the plane. Similar results are found when the magnetization is initially in-plane (not shown). Interestingly, four zones can be distinguished. For large positive in-plane torque, $\tau_{||}\geq \tau_{th}=\eta (1+Q_{||}-Q_{\bot})$ ($\eta$ is the Gilbert damping), the magnetization is stable in the perpendicular direction. When $\tau_{||}\leq \tau_{th}$, out-of-plane [OPP - Fig. \ref{fig:Fig4}(a)] and in-plane [IPP - Fig. \ref{fig:Fig4}(b)] magnetization precessions appear, while for even larger negative in-plane torque the in-plane anisotropy overcomes the perpendicular anisotropy and in-plane stable states are reached [see Fig. \ref{fig:Fig4}(c)]. The dependence of the resonance frequency as a function of the in-plane torque is reported on Fig. \ref{fig:Fig4}(d), where the positive slope (left hand) is attributed to out-of-plane precessions, whereas the negative slope (right hand) is attributed to in-plane precessions.\par

The present current-driven spin torque can be readily compared with the conventional STT \cite{slonc} and the SOI-ST \cite{manchon,soitorque}. First, the present torque has its origin in the interfacial SOI, rather than in the inhomogeneous or discontinuous magnetic texture so that no external polarizer is needed. Secondly, it possesses two components, $T_{||}$ and $T_{\bot}$, whereas the SOI-ST calculated in Ref. \cite{manchon,soitorque} produces only an effective magnetic field. Furthermore, similarly to the conventional STT, the in-plane torque $T_{||}$ competes with the damping and can excite self-sustained magnetic precessions in the absence of external magnetic field, provided that stable in-plane and out-of-plane magnetization stable state can be achieved.

For the parameters exploited here, the spin torque can be as large as 100eV/$\mu$ m$^2$ (see Fig. 2(b)), which is comparable to spin transfer torque in MTJs \cite{mtj}. From an experimental point of view, the key element needed to observe such spin torque is a large spin-orbit splitting at the interface between the ferromagnet and the barrier. As mentioned in the introduction, a number of results have been produced on SMTJs based on metals \cite{park} or semiconductors \cite{gould}. A good rule of thumb to design such efficient interfaces is to maximize the TAMR, as shown in Ref. \cite{park}. An efficient procedure to detect such a torque in the case of small bias (or small Rashba parameter) is to investigate spin torque-driven ferromagnetic resonance \cite{petit} or spin-diode effects \cite{spindiode}.

\begin{acknowledgments}
The author aknowledges fruitful discussions with S. Zhang, K.-J. Lee and P. J. Kelly. This work was supported by KAUST Academic Excellence Alliance funding program.
\end{acknowledgments}

\end{document}